\documentclass{optica-article}

\journal{opticajournal} 

\articletype{Research Article}

\usepackage{lineno}
\usepackage{bm}
\usepackage{subcaption}
\usepackage{siunitx}
\usepackage{xfrac}
\usepackage{physics}

\begin{document}
\title{Sub-quadratic scalable approximate linear converter using multi-plane light conversion with low-entropy mode mixers}

\author{Yoshitaka Taguchi,\authormark{1,*}}

\address{\authormark{1}Department of Electrical Engineering and Information Systems, The University of Tokyo, 7-3-1 Hongo, Bunkyo-ku, Tokyo 113-8656 Japan}

\email{\authormark{*}ytaguchi@ieee.org} 


\begin{abstract*} 
Optical computing is emerging as a promising platform for energy-efficient, high-throughput hardware in deep learning. A key challenge lies in the realization of optical matrix-vector multiplication, which often requires $O(N^2)$ modulators for exact synthesis of $N \times N$ matrices, limiting its scalability.
In this study, we propose an approximate matrix realization method using multi-plane light conversion (MPLC) that reduces both the system size and the number of phase shifters while maintaining acceptable error bounds.
This approach introduces low-entropy mode mixers, in which couplings between optical modes are weak. We demonstrate that such mixers can preserve computational accuracy while reducing hardware requirements, enabling more flexible and compact implementations.
We further investigate MPLC converters with fewer phase shifter layers than the theoretical minimum, and show that they function as approximate converters by tolerating predefined error thresholds, achieving sub-quadratic scaling.
To identify efficient architectures for implementing general linear matrices using unitary converters based on MPLC, we compare block-encoding (BE) and singular-value decomposition (SVD) schemes.
Results indicate that BE exhibits superior iterative configuration properties. By characterizing the trade-offs between entropy of mode mixers, number of phase shifters, and the error tolerance, this study provides a framework for designing scalable and efficient approximate optical converters.
Finally, using model quantization techniques, we demonstrate that the proposed method reduces the number of required phase shifters by half while preserving the accuracy of a text classification model.
\end{abstract*}

\section{Introduction}
\label{sec:intro}
Optical computing is emerging as a promising hardware platform for energy-efficient and high-throughput computing \cite{Shen2017,Zuo2019,Miscuglio2020,Ryou2021,Zhang2021,Rahman2023,Zhou2023,Xi2023,Xu2024,Bandyopadhyay2024}.
As modern deep learning models rely on architectures with a large number of parameters, scalability on photonic computing platforms, particularly in optical matrix vector multiplication, has gained attention \cite{Feldmann2021,Guo2022,Valensise2022,Nakajima2023}.
A major challenge in this area is the growing system size and the increasing number of components, which scale quadratically for the matrix multiplication device. 
Phase-shifter-based photonic circuits can achieve arbitrary unitary transformations as well as real-valued matrices, which is commonly used in deep learning models \cite{Shen2017,Fang2019,Tang2024,Fldzhyan2024}.
On integrated photonics platforms, cross-bar array wiring of thermo-optic phase shifters \cite{Antonio2020,Gurses2022} and phase-change-material based modulators \cite{Mario2020,Xuan2020,Miyatake2024} can reduce the number of wires required. While this approach achieves linear scaling in terms of wiring, the low operational speed of these modulators limits their applications. Alternatively, wavelength multiplexing parallelizes the calculation to reduce the system size \cite{Yang2012,Tohru2019,Frank2022}, but quadratic scaling challenges remain.
Another approach involves training models directly on photonic platforms with smaller number of components, embedding the physical system's properties into the model itself \cite{ZhangHui2021,Pegios2022,Cong2022,Saumil2022}. However, this strategy requires training models from scratch, or additional training of the pre-trained models developed on normal computers. This limits the reuse of existing machine learning assets.

On semiconductor-based computing platforms, model quantization is a technique developed in the area of computer science and is widely adopted for running the deep learning models efficiently. Quantization of model weights, which represents weight values in the model using fewer bits, reduces both the complexity of semiconductor circuits in computing units and memory requirements \cite{Courbariaux2015,Polino2018,Zafrir2019,Burgess2019}. During quantization, weights are converted to integer representations, introducing quantization errors into the model.
Extensive research has been conducted to address these errors, enabling quantized models to achieve performance comparable to their original counterparts. Remarkably, even models with weights represented using only a few bits can maintain its performance \cite{Jerry2023,Frantar2023,Xu2024Onebit}.
While these approximation techniques are widely implemented on semiconductor computing platforms, they have yet to be extensively studied in the field of optical computing. Given that optical computing is inherently analog, adopting proactive approximation methods can leverage its unique computational characteristics.

In this study, we propose a compact and scalable programmable linear converter based on multi-plane light conversion (MPLC) \cite{Morizur2010,Labroille2014,Tang2018} for approximate computation. The MPLC architecture has demonstrated its capability to implement target unitary transformations and real-valued linear conversions with high robustness to fabrication errors.
While prior research has established exact configuration methods for MPLC architectures \cite{Fontaine2019,Kuzmin2021,Bantysh23,Taguchi2023}, their potential for approximate computation remains unexplored. Here, we investigate the approximation capabilities of MPLC by analyzing the effects of varying the entropy of mode mixers and the number of phase-shifter layers. Shannon entropy is employed as a metric for mode mixing, providing a quantitative measure of how effectively a linear transformation mixes different modes.
Previous research has shown that the MPLC architecture using mode mixers with low entropy can be universally configurable \cite{Tanomura2021,TanomuraCLEOPR2022,Zelaya2024}; however, the quantitative relationship between matrix entropy and configurability remains unexplored. Our findings reveal that the relationship between entropy and matrix error after the configuration exhibits a plateau, indicating that MPLC architectures can be designed with low-entropy mode mixers without compromising error tolerance.
By utilizing mode mixers with low entropy, it is possible to employ compact mixers, thereby reducing the overall system size and adding design flexibility.
Furthermore, we demonstrate that allowing for some error in the realized matrix enables sub-quadratic scaling of the number of phase shifters, while conventional exact synthesis shows quadratic scaling.
We evaluate the approximation capability by measuring the maximum element-wise error in the synthesized matrix, offering an intuitive, quantization-like assessment of accuracy. The error bound is analyzed as a function of matrix entropy and the number of phase shifters, providing design guidelines for constructing approximate linear converters within a specified error tolerance.
Finally, our approximation approach is validated by applying to text classification model, where model quantization is found to be crucial for maintaining model accuracy under approximation.

This paper is organized as follows. In Section \ref{sec:intro_shannon_etpy}, we begin by discussing the fundamental properties of Shannon matrix entropy in relation to unitaries and mode mixing. We introduce a matrix interpolation between a given matrix and the identity matrix within the unitary group, providing a method for sampling unitary matrices with a specified Shannon entropy. We statistically investigate the relationship between entropy and mode mixing, offering an intuitive interpretation of Shannon entropy.
In Section \ref{sec:exact_low_entropy}, we examine the exact synthesis of unitary matrices and general matrices using the MPLC architecture. First, we demonstrate that even with low-entropy mode mixers, a few-redundant MPLC architecture can be configured to a target unitary matrix. The accuracy of matrix realization depends monotonically on the matrix entropy of the mode mixers. These results indicate that Shannon matrix entropy serves as a critical metric for designing mode mixers in MPLC architectures. Second, we compare two major synthesis schemes of general matrices and show that the block encoding (BE) scheme is advantageous over the singular value decomposition (SVD) scheme in terms of configuration efficiency and the number of required phase shifters. We also provide an algebraic proof of the minimum number of layers necessary for universal synthesis of matrices using the BE scheme.
Section \ref{sec:approx_converter} examines approximate converters based on MPLC architectures with an insufficient number of layers. While these converters cannot achieve exact synthesis, the number of required phase shifters scales sub-quadratically with the matrix size $N$, given a tolerable error bound. We systematically evaluate approximation errors by varying the number of phase-shifter layers and the matrix entropy.
In Section \ref{sec:model_quantization}, we train a text classification model and investigate the effect of approximation numerically.
By applying model quantization techniques developed in the field of deep learning, we demonstrate that the number of phase shifters required in the approximate converter can be reduced by half while maintaining model accuracy.
Section \ref{sec:conclusion} concludes this paper.

\section{Shannon entropy of unitary matrices}
\label{sec:intro_shannon_etpy}
In this section, we introduce Shannon entropy for mode mixers and outline its fundamental properties. Shannon entropy is defined for unistochastic matrices \cite{Karol2003}, which are widely studied in fields such as quantum mechanics on graphs \cite{Kottos1997,Gregor2001,Gnutzmann2005,Nechita2023}. A unistochastic matrix $B$ is derived from a unitary matrix $U$ as $B_{ij} = \abs{U_{ij}}^2$. For a complex vector $\bm{v}$, each element of $B$ represents the proportion of mixing among the components of $\bm{v}$ induced by the linear transformation $U\bm{v}$. The matrix $B$ satisfies the following conditions thanks to the unitarity of $U$:
\begin{equation}
	B_{ij} \geq 0, \quad \sum_{i} B_{ij}=1, \quad \sum_{j} B_{ij}=1,
\end{equation}
resembling the properties of a discrete probability distribution for the rows and columns of $B$.
Applying the entropy function $-\sum p(x) \ln p(x)$ for each row of $B$ defines the Shannon entropy for a unistochastic matrix $B$ \cite{Arasu2019}:
\begin{equation}
	-\frac{1}{N} \sum_{ij} B_{ij} \ln B_{ij},
\end{equation}
where $N$ is the matrix size.
The Shannon entropy ranges between its minimum value of $0$ and maximum value of $\ln N$ \cite{Bengtsson2005}.
In this study, we define the normalized Shannon entropy for a mode mixer with a transfer matrix represented by $U$ as:
\begin{equation}
	\mathcal{H}(U) = -\frac{1}{N \ln N} \sum_{ij} \abs{U_{ij}}^2 \ln \abs{U_{ij}}^2,
\end{equation}
where the normalization by $\ln N$ ensures that $0 \leq \mathcal{H}(U) \leq 1$.
Unitary matrices satisfying $\mathcal{H}(U) = 0$ correspond to permutation matrices, where each row and column contains only one nonzero entry equal to one.

\begin{figure}[htb]
\centerline{\includegraphics[width=85mm]{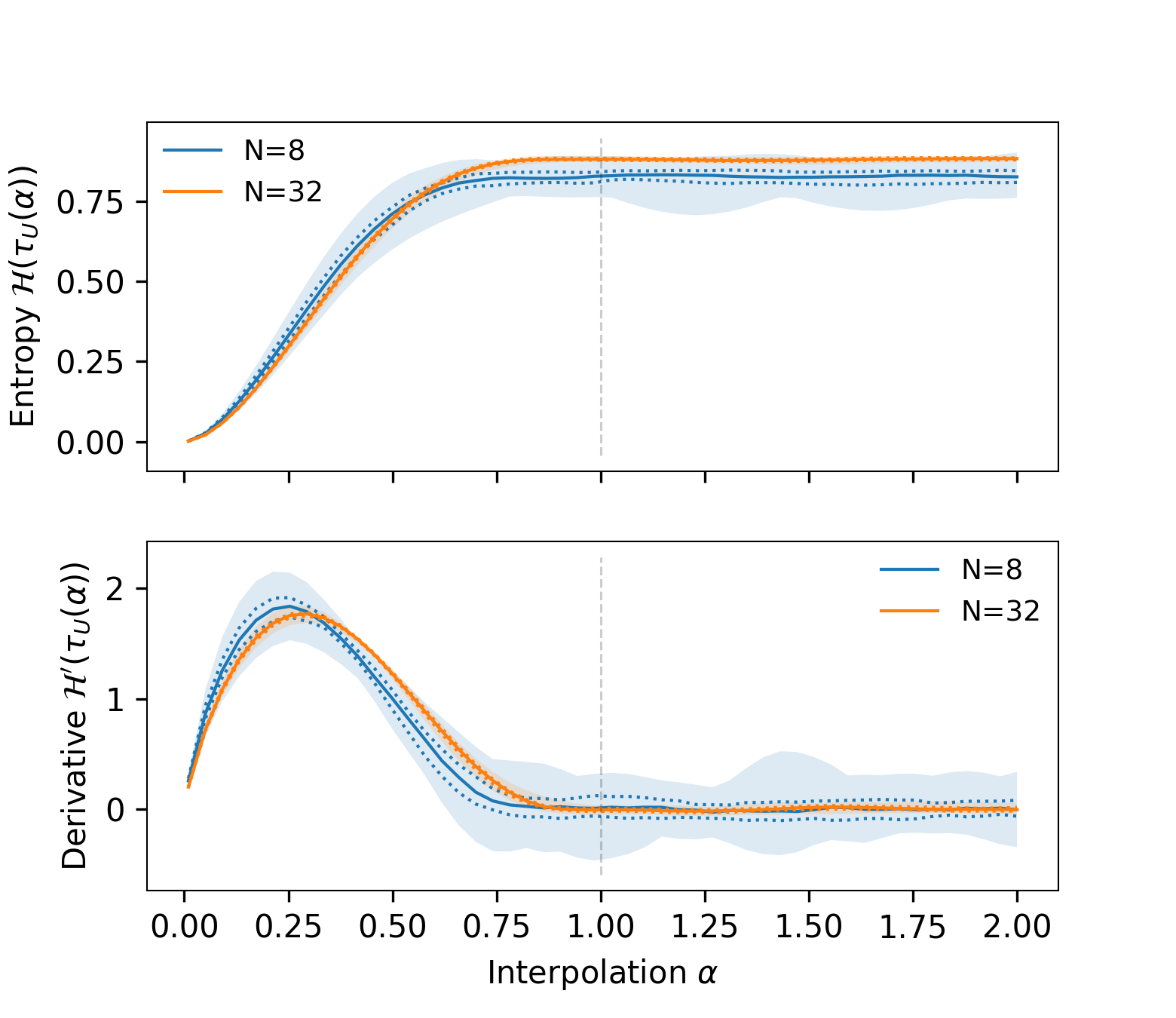}}
\caption{Shannon matrix entropy $\mathcal{H}(\tau_U(\alpha))$ and its derivative $\mathcal{H}'(\tau_U(\alpha))$ as functions of the interpolation parameter $\alpha$. The values are computed while sampling the unitary matrix $U$ for 64 times. The shaded area represents minimum and maximum values, the solid line represents the median, and the dotted line represents 25\% and 75\% quantiles of samples, respectively.}
\label{fig:entropy_interp}
\end{figure}

We define an interpolating function $\tau_U(\alpha): \mathbb{R}_{\geq 0} \to U(N)$ for a given unitary matrix $U$:
\begin{equation}
	\tau_U(\alpha) = U^\alpha = V^{-1} D^\alpha V,
\end{equation}
where $U = V^{-1} D V$ represents the diagonalization of $U$, and $D^\alpha = \mathrm{diag}({d_{11}}^\alpha, {d_{22}}^\alpha, \cdots {d_{nn}}^\alpha)$. 
For $0 \leq \alpha \leq 1$, $\tau_U(\alpha)$ interpolates on the unitary group between the identity matrix $I$ and the given matrix $U$, satisfying $\tau_U(0) = I, \tau_U(1) = U$, and $\tau_U(\alpha) \in U(N)$. 
We investigate the behavior of Shannon entropy under this interpolation $\mathcal{H}(\tau_U(\alpha))$ and demonstrate its statistical monotonicity. Figure \ref{fig:entropy_interp} presents $\mathcal{H}(\tau_U(\alpha))$ and its derivative $\mathcal{H}'(\tau_U(\alpha))$ as a function of $\alpha$ for $N=8$ and $N=32$. These values are evaluated for 64 samples of unitary matrices $U$.
The shaded area shows the range of minimum and maximum values, the dotted line shows the 25\% and 75\% quantiles, and the solid line shows the median.
The unitary matrix $U$ is sampled from Haar measure using the \texttt{stats} module of SciPy \cite{ScipyNmeth}.
The results show that the quantiles of Shannon entropy monotonically increase for $0 \leq \alpha \leq 1$ and saturates for $1 \leq \alpha$, especially for $N=32$. The variation of entropy and its derivative become smaller as $N$ increases. 
It is worth noting that similar monotonic behavior has been reported for von Neumann entanglement entropy in the time evolution of quantum systems \cite{Calabrese2005,Kazuya2017,Yohei2020}.
This monotonic property is utilized throughout this paper to sample unitary matrices with a specified target entropy, i.e., sampling $U$ such that $\mathcal{H}(U) = t$ for given $t$.

\begin{figure}[htb]
\centerline{\includegraphics[width=82mm]{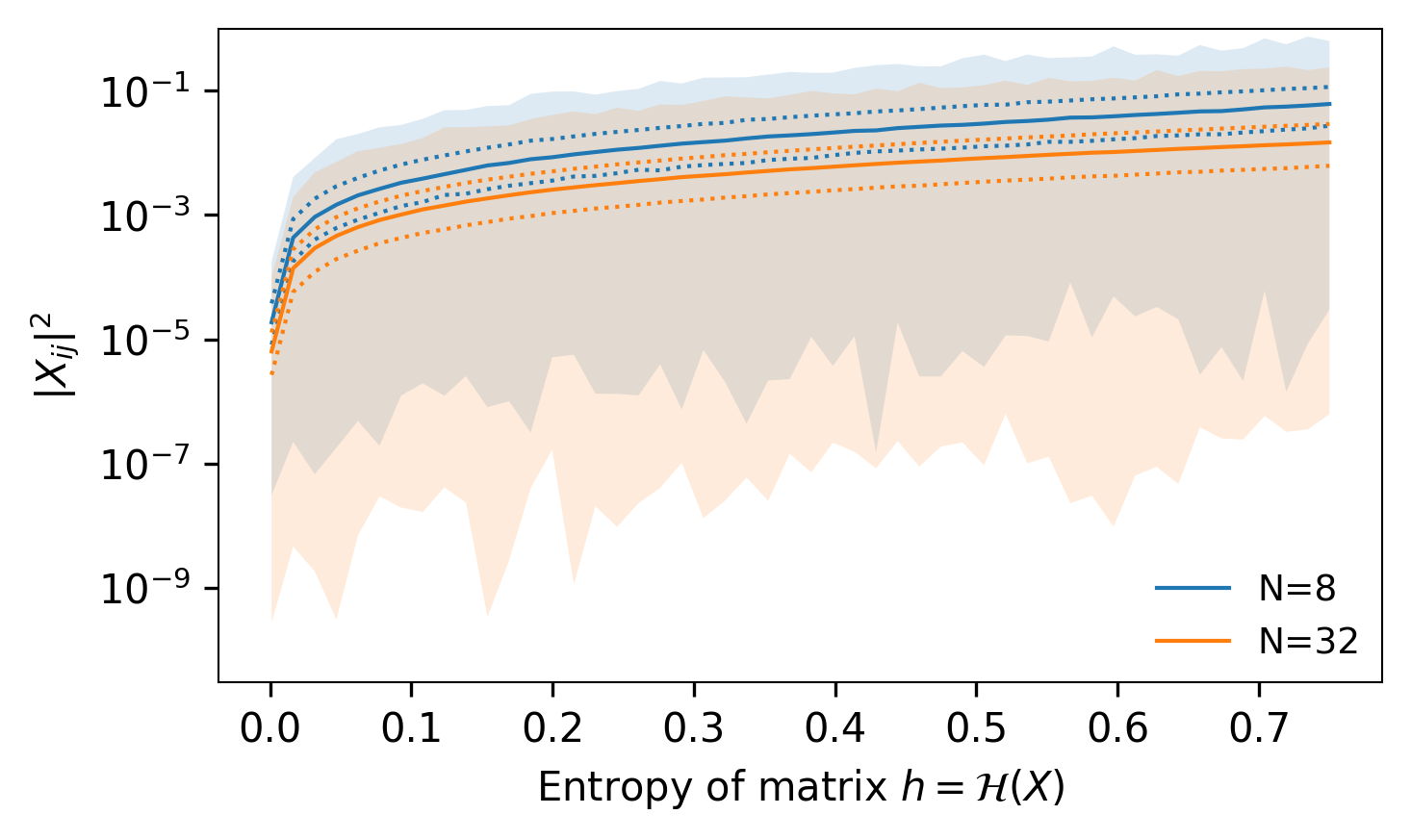}}
\caption{Squared absolute values of non-diagonal elements for unitary matrices, $\abs{X_{ij}}^2$, with $i \neq j$, having a specified entropy $h = \mathcal{H}(X)$. The 64 samples are shown in the same manner as Figure \ref{fig:entropy_interp}.}
\label{fig:stat_nondiagonal}
\end{figure}

We further examine the relationship between the non-diagonal matrix elements and entropy. Figure \ref{fig:stat_nondiagonal} illustrates the range, quantiles, and median of the squared absolute values of non-diagonal elements, $\abs{X_{ij}}^2$, for $i \neq j$, as a function of entropy $h$.
For a given target entropy $h$, matrix $X$ satisfying $\mathcal{H}(X) = h$ is sampled 64 times, and the corresponding statistics are computed.
The sampling process involves numerically solving the equation $\mathcal{H}(\tau_U(\alpha)) = h$ for $0 \leq \alpha \leq 1$, following the initial sampling of a unitary matrix $U$. 
This equation always yields a unique solution due to the monotonic relationship shown in Figure \ref{fig:entropy_interp}. 
The quantiles of the non-diagonal elements exhibit a monotonic relationship with the Shannon entropy, confirming that entropy effectively measures the degree of mixing in a given unitary matrix.

From the device design perspective, it is worth noting that mode mixers having a specific normalized Shannon entropy always exist. The existence follows from the continuity of Shannon entropy and the existence of equal-splitting mixers. Shannon entropy is maximized to $1$ when the power from each mode in the mode mixer is equally split. Such design exists for arbitrary $N$. On the contrary, Shannon entropy is minimized to $0$ when the mixer does not mix the modes at all.
Because Shannon entropy is defined as a continuous function of the elements of matrix, any continuous design parameter varies the Shannon entropy continuously.
Therefore, intermediate value theorem indicates that there always exists a mixer that has an arbitrary Shannon entropy. While we can show the existence of mixers, the direct design methodology of mixers from a given Shannon entropy remains unclear and further discussion is necessary. Nevertheless, if some parameter of mixer, such as coupling length, monotonically increases the Shannon entropy, we can utilize binary search to efficiently determine the parameter that achieves the target Shannon entropy.

\section{Exact converter with low-entropy mode mixers}
\label{sec:exact_low_entropy}
In this section, we explore the relationship between the Shannon entropy of mode mixers in the MPLC architecture and its exact configuration capabilities.
Previous studies have examined the requirements for mode mixers to enable universal configuration in MPLC architectures using specific mode mixers \cite{Tanomura2022,TanomuraCLEOPR2022,Markowitz2023}.
Here, we focus on the configuration capabilities of MPLC architectures with mode mixers characterized by their Shannon entropy. Both the synthesis of unitary matrices and general linear matrices are analyzed. For the synthesis of general linear matrices, we compare two approaches: the SVD scheme and the BE scheme. Our results demonstrate that the entropy of mode mixers can be reduced without compromising the universality, and that the BE scheme is advantageous for iterative configuration.

\subsection{Device definition}
\begin{figure}[hbtp]
\centerline{\includegraphics[width=75mm]{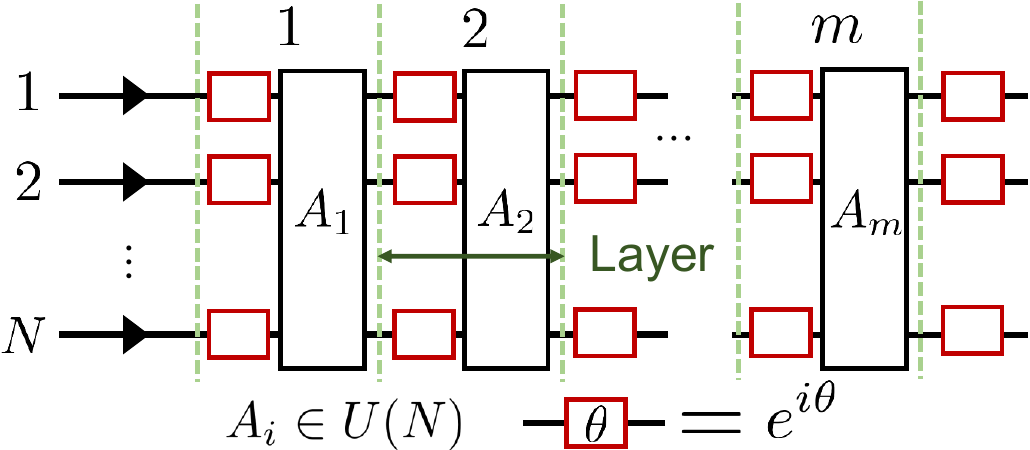}}
\caption{Schematic representation of the $N\times m$ MPLC architecture. The number of layers is specified by $m$. Each layer consists of an $N$-port fixed unitary converter represented by $A_i$, followed by an array of $N$ single-mode phase shifters.}
\label{fig:MPLC}
\end{figure}

We define the unitary converter based on the MPLC architecture.
In this paper, we use the few-layer redundant MPLC architecture \cite{Taguchi2023,Taguchi2023CLEO,Markowitz2023}. This architecture has demonstrated its ability to be exactly configured to any target unitary using an iterative optimization method.
The structure of this architecture is illustrated in Figure \ref{fig:MPLC}.
Each layer of the architecture consists of an $N$-port fixed unitary mode mixer $A_i$ and an array of $N$ single-mode phase shifters. The overall transformation of this device, denoted as $X$, is given by
\begin{equation}
X = L_{m+1} A_m L_m \cdots A_2 L_2 A_1 L_1,
\end{equation}
where $A_i$ is the transfer matrix of a $N$-port unitary mode mixer, and $L_k$ is defined as $L_k = \mathrm{diag}(e^{i\theta_{k1}}, e^{i\theta_{k2}}, \cdots, e^{i\theta_{kn}})$.
In the few-layer redundant configuration, the number of layers, $m$, is set to $N+1$ \cite{Taguchi2023}.
Each distinct phase shift within the phase shifters is denoted by a real parameter variable, and all the phase shifts are collectively expressed as a vector $\bm{p}$.

To evaluate the exact and universal synthesis capability of this architecture, we fix the Shannon entropy of all mode mixers $A_i$ to the same value. 
This uniform entropy setting is motivated by two considerations:
\begin{enumerate}
  \item If the entropy of all mixers is zero (i.e., the mixers do not perform any mode mixing), the unitary converter cannot achieve universality.
  \item Previous research \cite{Saygin2020,Tanomura2023} has demonstrated the robustness of the MPLC architecture to imperfections in its mode mixers. Based on this robustness, if a converter is universally configurable, increasing the entropy of one mixer should not compromise the overall universality of the converter.
\end{enumerate}
Therefore, by fixing the entropy equally across all mixers, we aim to determine the minimum entropy required for the architecture to be universal.

\subsection{Optimization problem setting and algorithm}
\label{sec:opt_setting}
The phase shifter optimization problem of exact unitary converter is formulated as follows. The normalized cost function $\mathcal{L}$ \cite{Taguchi2023} between two matrices is defined as:
\begin{equation}
\mathcal{L}(\bm{p}) = \frac{1}{4N} \norm{X(\bm{p}) - U}_F^2,
\label{eq:def_L}
\end{equation}
where $X({\bm{p}})$ is the unitary matrix realized physically by the parameter vector $\bm{p}$, $U$ is the target matrix to be achieved, and $\norm{\cdot}_F$ is the Frobenius norm.
To minimize $\mathcal{L}(\bm{p})$, we employ numerical optimization. At the start of the optimization, parameters are initialized using a uniform distribution ranging from $0$ to $2\pi$, and the target unitary matrix $U$ is sampled from the Haar measure using the \texttt{stats} module of SciPy \cite{ScipyNmeth}. In the MPLC architecture, each matrix $A_i$ for $1 \leq i \leq m$ is sampled so that the entropy of every mixer is fixed to $h$, that is, $\mathcal{H}(A_i) = h$.
This entropy-fixing sampling follows the procedure outlined in Section \ref{sec:intro_shannon_etpy}.
After initializing the parameters and the matrix, the cost function $\mathcal{L}$ is optimized using the quasi-Newton optimization method, specifically, the limited-memory Broyden-Fletcher-Goldfarb-Shanno (L-BFGS) algorithm \cite{Flet1987} implemented in \texttt{optimize} module of SciPy \cite{ScipyNmeth}.
This method starts from the initial parameters and modifies them at each step until convergence to the local minimum, where $d\mathcal{L}/d\bm{p}=\vb{0}$. The optimization is run 64 times while changing the initial parameters to investigate the statistical behavior.
Matrices $U$ and $A_i$ are sampled at the each optimization. The gradient, which is essential for L-BFGS optimization algorithm, is provided by the standalone gradient method \cite{Taguchi2024}. During the optimization, only information from the input and output vectors is used, and no explicit knowledge of the matrices $A_i$ is incorporated.

\subsection{Synthesis of general linear using unitaries}
\label{sec:syntheis_exact_gen}
To support broader applications, particularly in deep learning, matrix multiplication devices must be capable of handling not only unitary matrices but also general matrices. We assume that the matrix to be synthesized is scaled such that all its singular values are less than 1, and we refer to this matrix as a general linear matrix in this study. Two schemes for implementing general matrices using unitary matrices are considered: SVD and BE.
The SVD scheme realizes a general matrix by combining a layer of intensity modulators with two unitary converters \cite{Shen2017,Fang2019}. Specifically, the matrix is decomposed into two unitary matrices and a diagonal matrix, with the diagonal matrix implemented using the intensity modulator array.

The BE scheme, on the other hand, embeds the $N \times N$ matrix into a $2N \times 2N$ unitary matrix \cite{Andras2019,Daan2024,Tang2024,Fldzhyan2024}.
\begin{figure}[htbp]
\begin{minipage}[b]{0.95\linewidth}
	\centering
	\includegraphics[width=7cm]{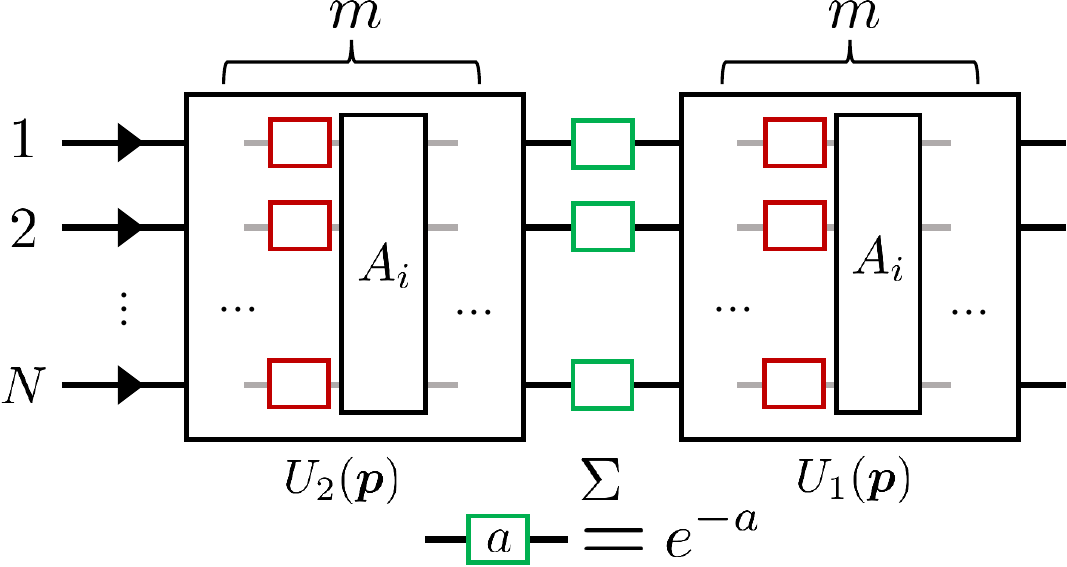}
	\subcaption{}
	\label{fig:schem_svdwmplc}
  \end{minipage}
  \begin{minipage}[b]{0.95\linewidth}
	\centering
	\vspace{0.5cm}
	\includegraphics[width=7cm]{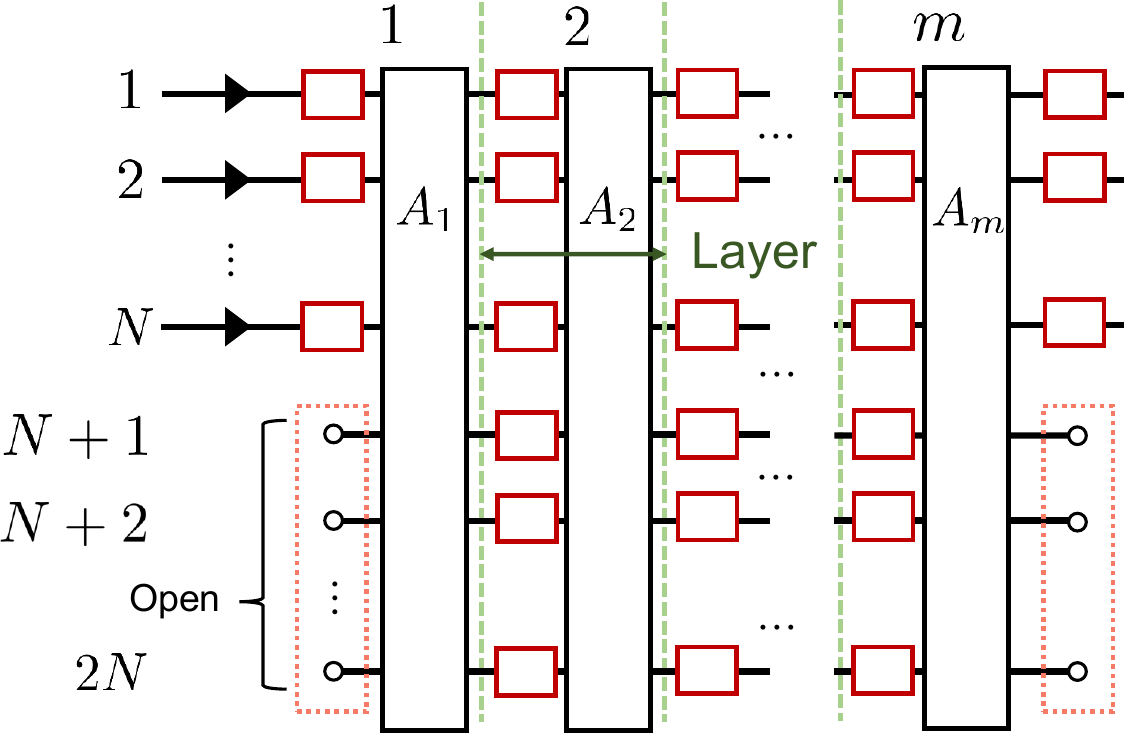}
	\subcaption{}
	\label{fig:schem_mplc}
\end{minipage}
  \caption{Schematics of the (a) SVD and (b) BE schemes.
  In the SVD scheme, the structure consists of a series connection of a unitary converter with $m$ layers, an array of intensity modulators denoted by $a$, and another unitary converter with $m$ layers.
  For the BE scheme, the general linear matrix is embedded into a $2N \times 2N$ unitary matrix using an MPLC unitary converter with $m$ layers. In this scheme, $N$ ports are open at the input and output of the mode mixers $A_1$ and $A_m$, respectively, out of the $2N$ ports.
 }
\label{fig:schem_unitary_converters}
\end{figure}

We define the cost function for configuring a general matrix and describe how the parameter vector $\bm{p}$ is mapped for the SVD and BE schemes. The cost function for general linear matrices is based on the Frobenius norm:
\begin{equation}
\mathcal{L}_\mathrm{gen}(\bm{p}) = \norm{Q(\bm{p}) - P}_F^2,
\label{eq:def_L_gen}
\end{equation}
where $Q(\bm{p})$ is the general linear matrix physically realized by the parameter vector $\bm{p}$, and $P$ is the target matrix to be achieved.
For numerical simulation, the target matrix $P$ is generated such that its singular values are less than 1, using $P=W_1\: \mathrm{diag}(\sigma_1, \sigma_2, \cdots, \sigma_n)\:W_2$, where $W_1$ and $W_2$ are Haar-random unitary matrices, and the $N$ singular values $\sigma_i$ are sampled from the uniform distribution $U(0, 1)$.

In the SVD scheme, $Q(\bm{p})$ is realized as:
\begin{equation}
	Q(\bm{p}) = X_1(\bm{p}) \; \Sigma \; X_2(\bm{p})
\end{equation}
where $X_1(\bm{p})$ and $X_2(\bm{p})$ are unitary matrices, and $\Sigma$ is a real diagonal matrix with elements less than 1, as illustrated in Figure \ref{fig:schem_svdwmplc}.
Half of the components in $\bm{p}$ are used to parameterize $X_1$, and the other half to parameterize $X_2$.
Here, we do not include the diagonal elements of $\Sigma$ in $\bm{p}$.
This is because the total optical power at the outputs depends only on the sum of the diagonal elements of $\Sigma$, allowing each intensity modulator to be configured separately by measuring the total output power and adjusting the corresponding intensity modulators.
Once the intensity modulators are set, the unitary converters, $X_1$ and $X_2$, are configured accordingly.
The standalone gradient method is used to compute the gradient $d\mathcal{L}_\mathrm{gen}/d\bm{p}$, which is essential for the optimization algorithm.

In contrast, the BE scheme does not use intensity modulators, and realizes a general $N \times N$ matrix $Q$ via the following embedding:
\begin{equation}
\label{eq:block_encoding_embedding}
Y = 
\begin{bmatrix}
	Q & \sqrt{I - Q Q^\dagger} \\[4pt]
	\sqrt{I - Q^\dagger Q} & -Q^\dagger
\end{bmatrix},
\end{equation}
where $Y$ is a $2N \times 2N$ unitary matrix, $I$ is the $N \times N$ identity matrix, and $\sqrt{\cdot}$ is the matrix square root. The $2N \times 2N$ matrix $Y$ is synthesized using the MPLC architecture, 
with half of the input and output ports employed to realize $Q$, as illustrated in Figure \ref{fig:schem_mplc}.
To determine the minimum number of layers $m$ required to synthesize $Y$ using the MPLC architecture, we first consider the degrees of freedom in the architecture.
The $2N \times 2N$ unitary matrix $Y$ has $(2N)^2 = 4N^2$ degrees of freedom, while the matrix square roots in the embedding Eq. \ref{eq:block_encoding_embedding} introduce additional unitary degree of freedom. A proof for the degree of freedom of the matrix square root is provided in the Appendix \ref{sec:proof_dof_root}.
Since any unitary can be chosen for the matrix square root, the embedding reduces the degrees of freedom of $Y$ to $4N^2 - 2N^2 = 2N^2$.
Considering that only half of the input and output ports are used, we can ignore half the phase shifters in the first and last layers of the MPLC architecture. 
Denoting the number of internal layers with $2N$ phase shifters by $m'$, the following inequality must hold to ensure the synthesis of $Q$ from the perspective of degrees of freedom:
\begin{equation}
\label{eq:block_encoding_least_layer_ineq}
(N-1) + m'(2N - 1) + (N-1) + 1 \geq 2N^2.
\end{equation}
Note that each phase shifter array has $N-1$ or $2N-1$ degrees of freedom due to the loss of one degree of freedom from the global phase, and the entire device has an additional degree of freedom corresponding to the global phase.
Solving this inequality, noting that $m'$ is an integer, leads to $m' > N - 1$.
Including the first and last layers, the total number of phase shifter layers satisfies $m' + 2 > N + 1$, establishing that the minimum number of phase-shifter layers required is $N+2$.
We now conclude that the minimum number of phase-shifter layers is $N + 2$.
This corresponds to $m=N+1$, consistent with previous work \cite{Tang2024}. While the proof for $m=N+1$ in that work 
relies on specific constructions of the matrix square roots, we provide a more general algebraic derivation.
For the optimization, each phase shifters are collectively represented by $\bm{p}$, and the matrix $Q(\bm{p})$ is obtained as the upper-left $N \times N$ submatrix of $Y(\bm{p})$.
Standalone gradient method \cite{Taguchi2024} is used to compute the exact gradient $d\mathcal{L}_\mathrm{gen}/d\bm{p}$ for this submatrix.
Similarly the optimization process for synthesizing unitaries in Section \ref{sec:opt_setting}, only information from the input and output ports is utilized during optimization.

\subsection{Results}
\label{sec:result_exact_synthesis}
\begin{figure}[hbtp]
\centerline{\includegraphics[width=75mm]{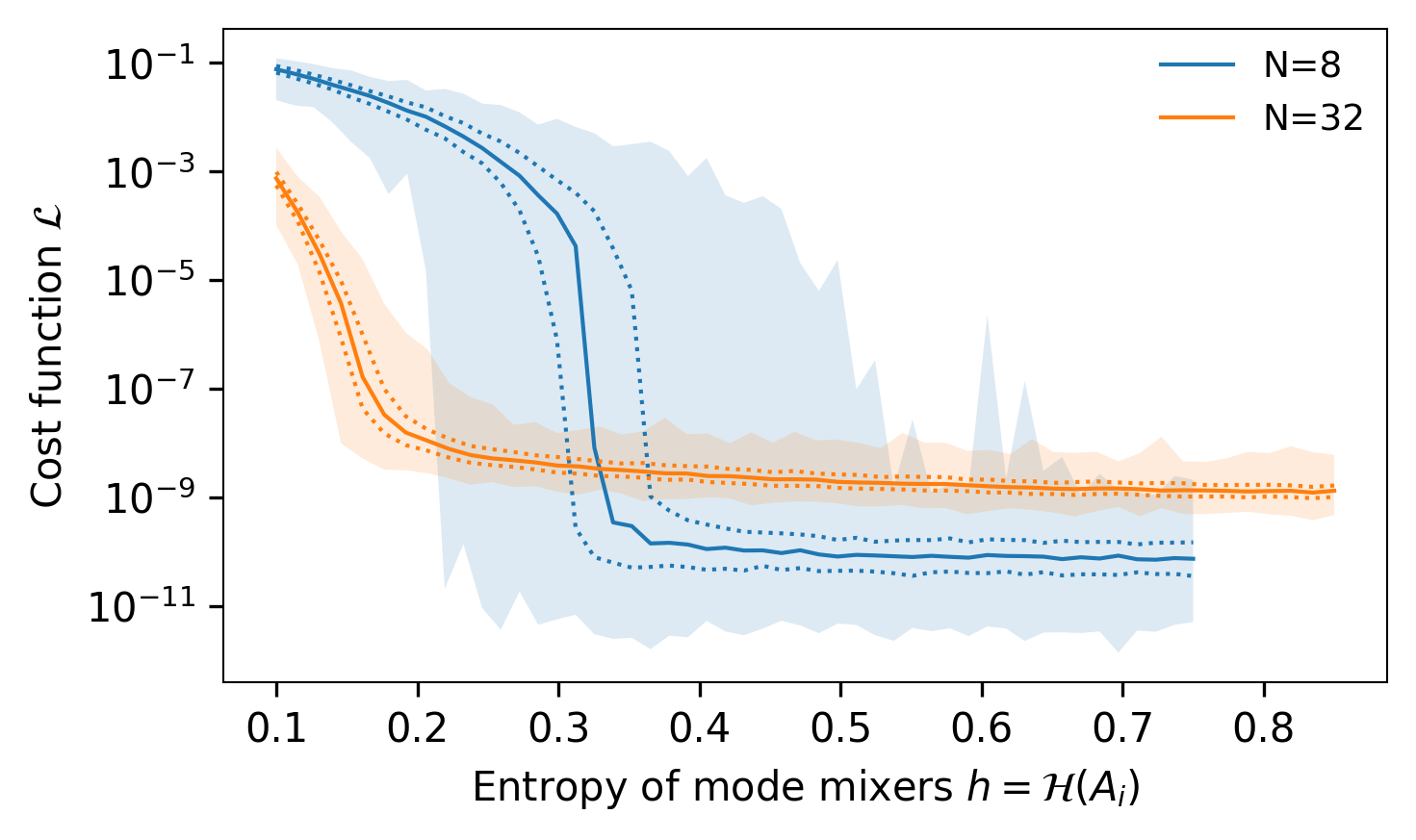}}
\caption{
Cost function $\mathcal{L}$ as a function of entropy $h$ for $N=8$ and $N=32$, where all mixers $A_i$ satisfy $\mathcal{H}(A_i) = h$. Results from 512 optimization are shown in the same manner as in Figure \ref{fig:entropy_interp}.
}
\label{fig:entropy_vs_frobenius_by_unitary}
\end{figure}

Figure \ref{fig:entropy_vs_frobenius_by_unitary} shows the cost function $\mathcal{L}$ after optimization as a function of the entropy of mode mixers for the exact unitary converter configuration. 
The median of the cost function is recorded 512 times after optimization while sampling the target matrix $P$ and mixers $A_i$.
For both cases, $N=8$ and $N=32$, the cost function decreases as the entropy of the mode mixers increases. Moreover, the reduction in the cost function saturates when the entropy exceeds a certain value. This indicates that mode mixers with lower entropy can be used without compromising universality, enabling the use of more compact mixers and a wider range of designs.
For instance, in the $N=32$ case, mode mixers with an entropy of $h = 0.3$ can be used without increasing the cost function $\mathcal{L}$ significantly. At this entropy, at least 75\% of the couplings among modes in the mixers are below 1\%, as shown in Figure \ref{fig:stat_nondiagonal}.

\begin{figure}[hbtp]
\begin{subfigure}{0.5\textwidth}
\includegraphics[width=\textwidth]{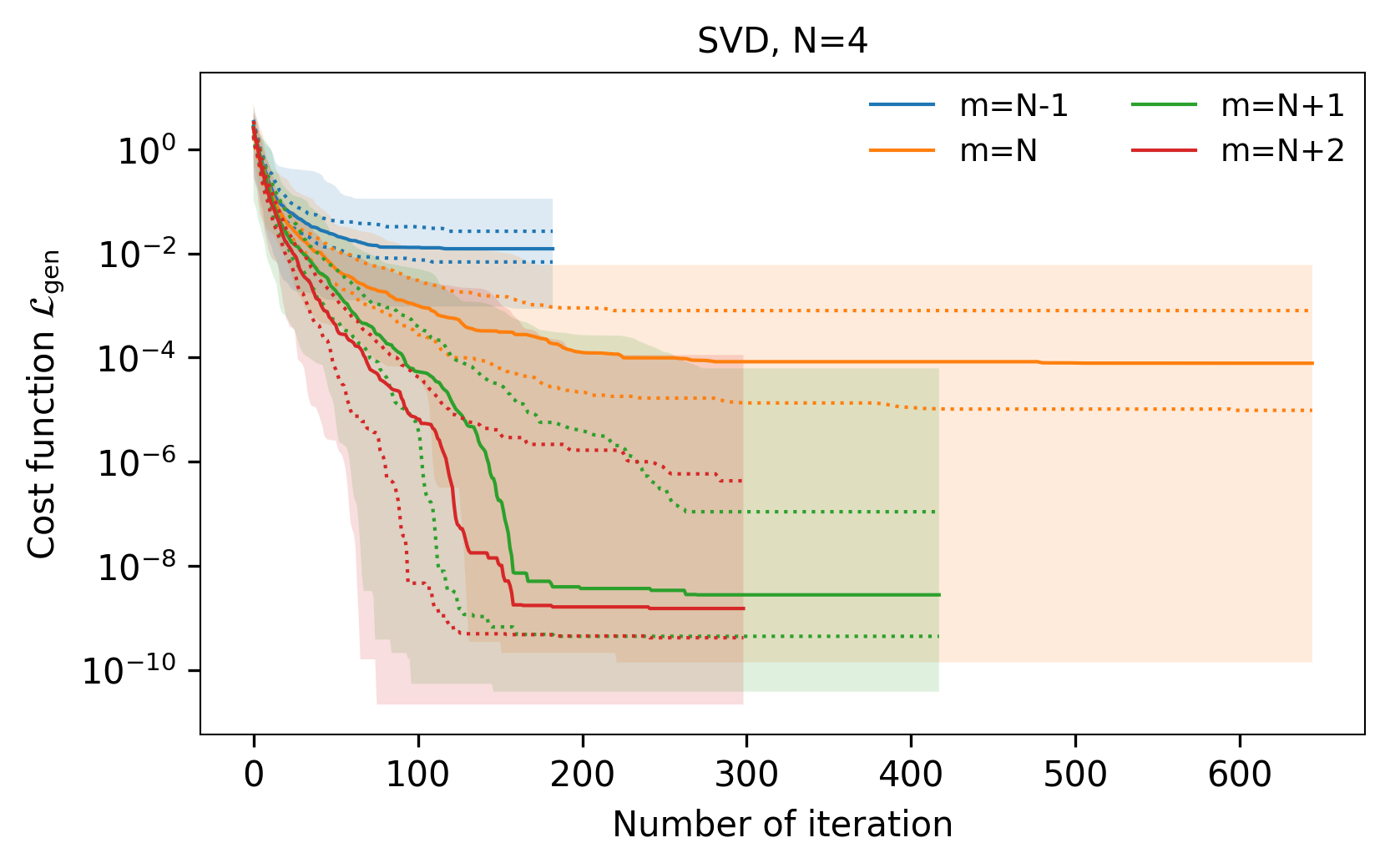}
\caption{}
\label{fig:embed_comp_SVDN4}
\end{subfigure}
\begin{subfigure}{0.5\textwidth}
\includegraphics[width=\textwidth]{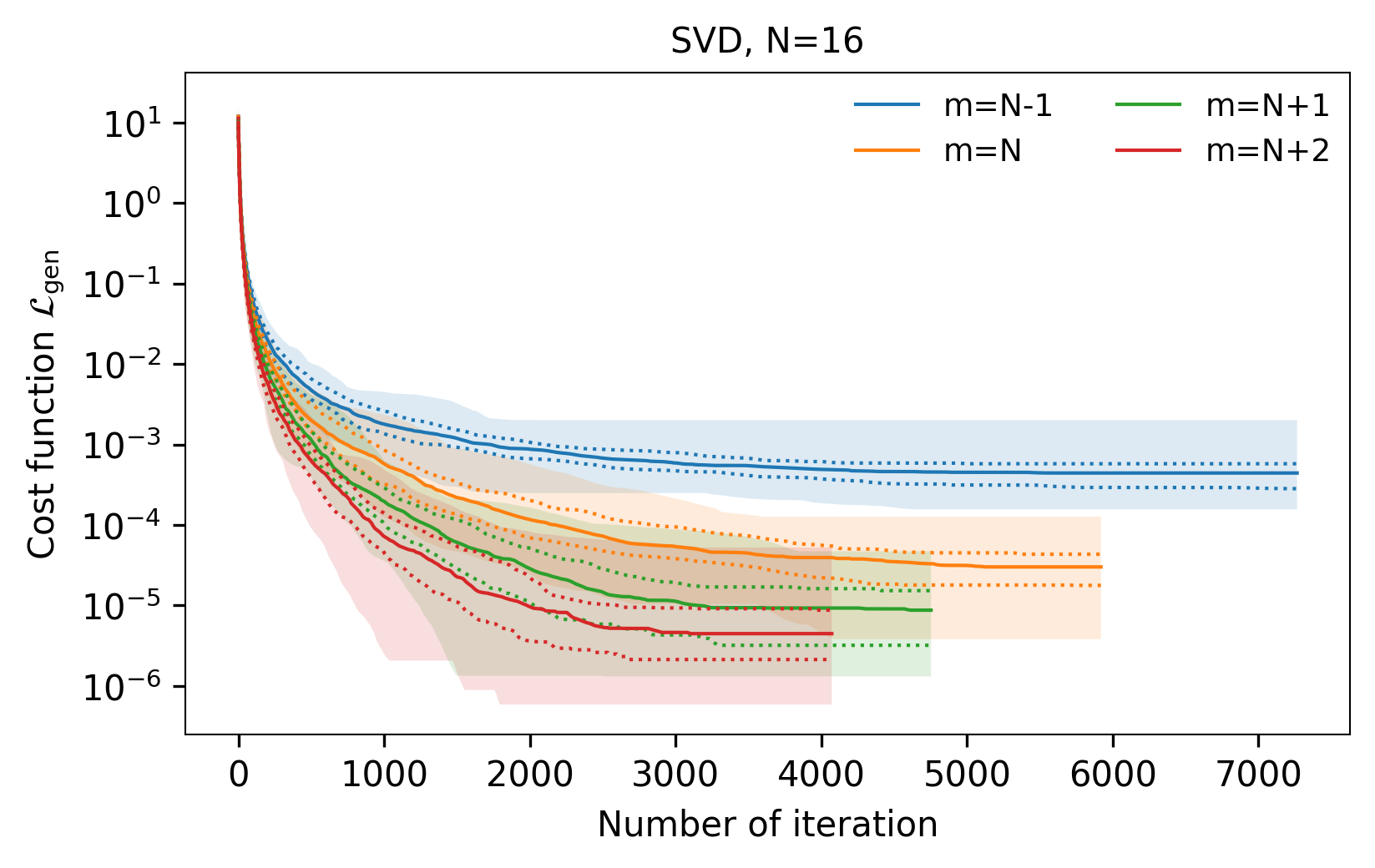}
\caption{}
\label{fig:embed_comp_SVDN16}
\end{subfigure}
\begin{subfigure}{0.5\textwidth}
\includegraphics[width=\textwidth]{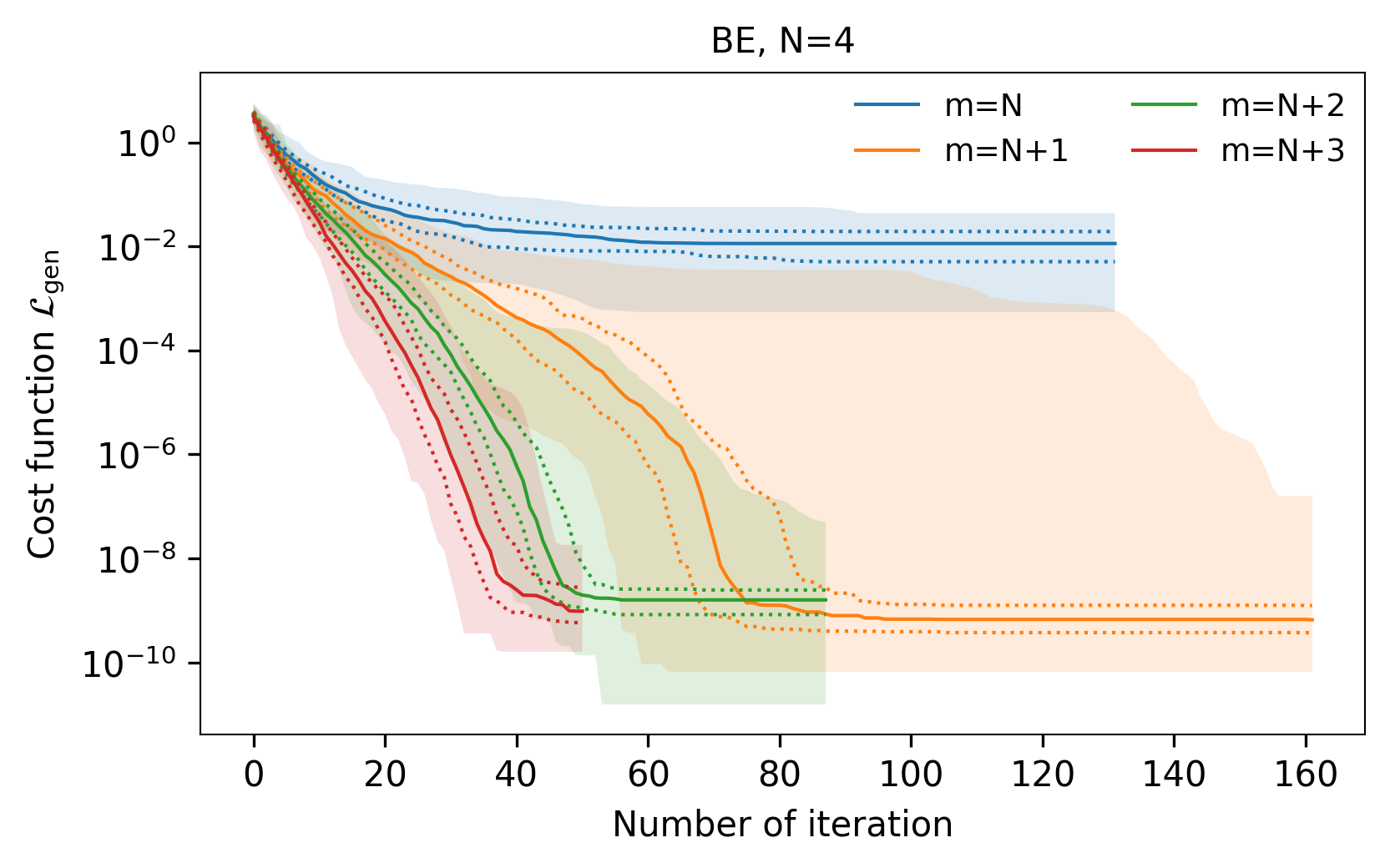}
\caption{}
\label{fig:embed_comp_BEN4}
\end{subfigure}
\begin{subfigure}{0.5\textwidth}
\includegraphics[width=\textwidth]{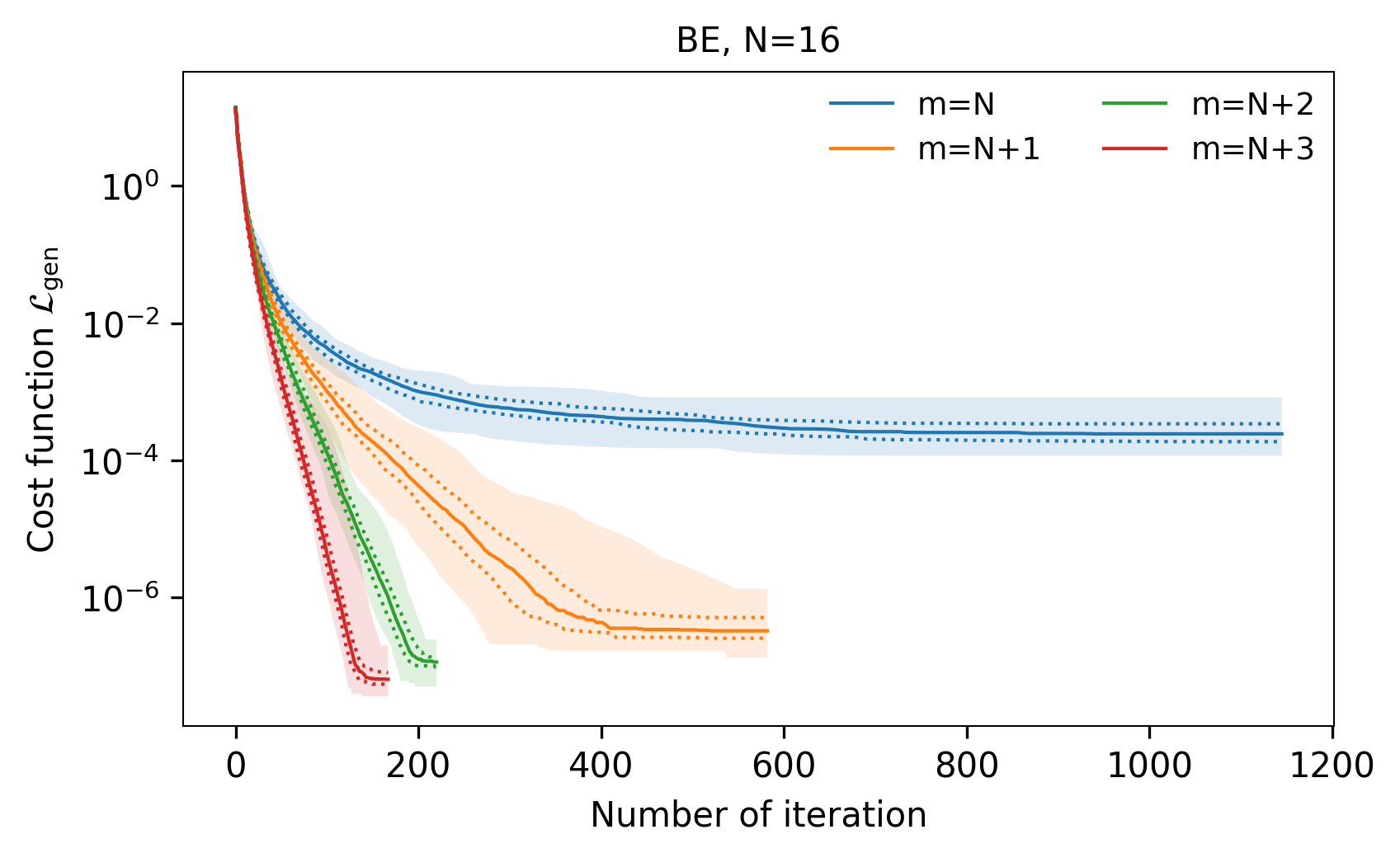}
\caption{}
\label{fig:embed_comp_BEN16}
\end{subfigure}
\caption{Convergence plots for SVD and BE schemes. The vertical axis shows the value of the cost function $\mathcal{L}_\mathrm{gen}$ defined by Eq. \ref{eq:def_L_gen}, and the horizontal axis shows the number of iterations.
The shaded area represents the minimum and maximum values, the solid line represents the median, and the dotted line represents the 25\% and 75\% quantiles over 64 optimization trials.
SVD with (a) $N=4$, (b) $N=8$, BE with (c) $N=4$, and (d) $N=16$.}
\label{fig:embed_comp}
\end{figure}

Figure \ref{fig:embed_comp} shows the convergence plots comparing the SVD and BE schemes for the configuration of a general linear converter. In the SVD scheme, $m=N$ represents the minimum number of layers required to achieve the $N^2$ degrees of freedom in an $N \times N$ unitary matrix.
A significant change is observed when increasing from $m=N$ to $m=N+1$ in both the number of iterations required for convergence and the final value of the cost function $\mathcal{L}_\mathrm{gen}$, as shown in Figure \ref{fig:embed_comp_SVDN4} and Figure \ref{fig:embed_comp_SVDN16}. This occurs because $m=N+1$ corresponds to the few-redundant configuration of MPLC in each unitary converter.
For BE scheme, $m=N+1$ is the minimum number of layers needed to provide the necessary degrees of freedom in the architecture. A substantial reduction in the cost function is observed when increasing from $m=N$ to $m = N+1$, as shown in Figs. \ref{fig:embed_comp_BEN4} and \ref{fig:embed_comp_BEN16}. While adding more layers reduces the number of iterations, it has only a minor impact on the final cost function value.
Overall, the BE scheme requires fewer iterations to converge compared to the SVD scheme and achieves a smaller final value of the cost function $\mathcal{L}_\mathrm{gen}$.
In contrast, the SVD scheme exhibits a large variance in the cost function across all cases, even though it has the sufficient number of degrees of freedom.

\begin{figure}[htbp]
\centerline{\includegraphics[width=75mm]{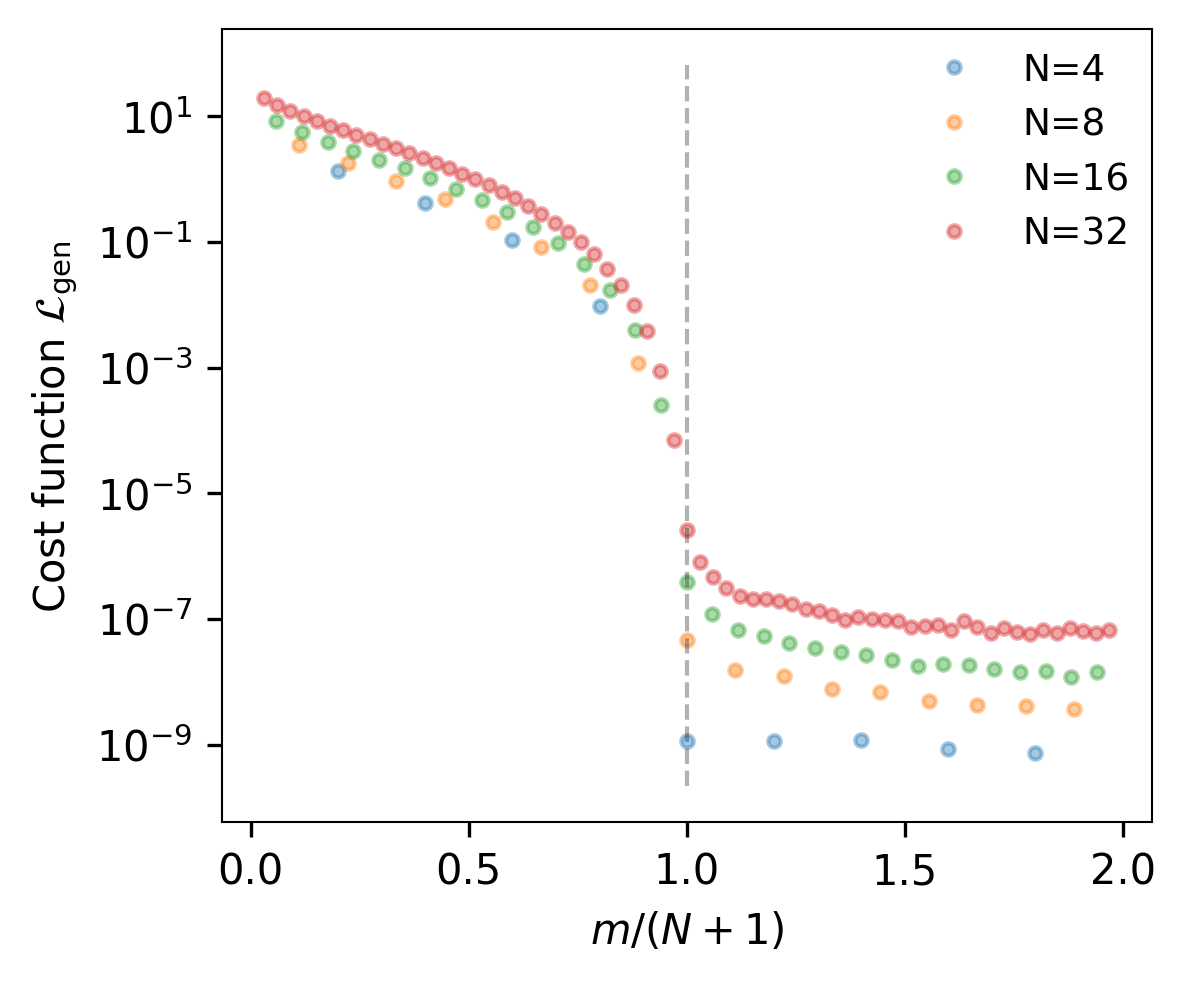}}
\caption{Median value of the cost function $\mathcal{L}_\mathrm{gen}$ after 64 optimization trials, shows as a function of $m / (N+1)$ for $N = 4, 8, 16, 32$.}
\label{fig:block_encoding_layer_vs_error}
\end{figure}

We further examine how the final value of the cost function changes with the number of layers in the BE scheme. 
Figure \ref{fig:block_encoding_layer_vs_error} shows the median of the final cost function $\mathcal{L}_\mathrm{gen}$ as a function of $m / (N+1)$, where the median is calculated across 64 optimization trials.
The number of layers $m$ is varied from $m=1$ to $m=2N$. For all cases ($N=4, 8, 16, 32$), a substantial decrease in the cost function is observed at $m / (N+1) = 1$, corresponding to $m = N+1$.
This indicates that redundant layers are not required to achieve a significant reduction in the cost function. This observation is consistent with the fact that Criterion~(\ref{eq:block_encoding_least_layer_ineq}) does not hold as an equality for integer values of $m'$, and that the condition $m = N + 1$ inherently introduces redundancy into the system. Such redundancy has been shown to improve convergence in MPLC architectures \cite{Taguchi2023}.

\section{Approximate converter with insufficient layers}
\label{sec:approx_converter}
In this section, we investigate the approximation abilities of the MPLC architecture when the number of layers is insufficient.
As shown in Figure \ref{fig:block_encoding_layer_vs_error}, the MPLC architecture retains error after optimization when the number of layers is less than the minimum required ($m < N+1$).
This observation suggests that if some error in the converter can be tolerated, the number of layers in the MPLC architecture can be reduced. Our results demonstrate that for a given upper bound on the error in matrix components, the required number of phase shifters in the approximate converter scales sub-quadratically. We adopt the BE scheme for the the following discussion, as it shows better convergence performance.

To measure the error in the approximate converter, we introduce an entry-wise maximum matrix norm defined as follows:
\begin{equation}
\norm{J}_\mathrm{max} = \max_{\substack{i, j = 1, \ldots, N}} \abs{J_{ij}},
\label{eq:def_entwise_max_norm}
\end{equation}
where $J$ is a matrix.
This definition is equivalent to the $L^\infty$ norm of a vector, $\norm{\vb{x}}_\infty = \underset{i}{\max} \abs{x_i}$, where all the matrix components are flattened and treated as elements of a vector. Since $\norm{\vb{x}}_\infty$ satisfies the conditions for a norm, the matrix norm defined in Eq. \ref{eq:def_entwise_max_norm} also satisfies the norm conditions. 
This norm represents the maximum absolute value of the matrix components, and $\norm{Q(\bm{p}) - P}_\mathrm{max}$ represents the maximum absolute error between the target matrix $P$ and the realized matrix $Q(\bm{p})$. By introducing this norm, we can analyze the approximation ability of the matrix in the context of quantization techniques used in deep learning models.

\subsection{Optimization problem setting and evaluation}
Instead of directly minimizing $\norm{\cdot}_\mathrm{max}$, we use the cost function $\mathcal{L}_\mathrm{gen}$ defined in Eq. \ref{eq:def_L_gen} for the optimization of the approximate converter. This is motivated by the fact that the derivative of $\norm{\cdot}_\mathrm{max}$ cannot be explicitly expressed and is not efficiently minimized, as no convenient formulation showing its convexity is known to the best of our knowledge. 
However, since $\mathcal{L}_\mathrm{gen}(\bm{p}) = 0 \Leftrightarrow \norm{Q(\bm{p}) - P}_\mathrm{max} = 0$, minimizing $\mathcal{L}_\mathrm{gen}(\bm{p})$ is expected to also minimize $\norm{Q(\bm{p}) - P}_\mathrm{max}$. Therefore, the same cost function is used as in the exact configuration.
For the approximate realization of an $N \times N$ matrix, the MPLC architecture with $2N$ ports and the BE scheme are employed. The optimization algorithm is same as that used for the exact synthesis with the BE scheme, as described in Section \ref{sec:syntheis_exact_gen}.

\begin{figure}[htbp]
\begin{subfigure}{0.33\textwidth}
\includegraphics[width=\linewidth]{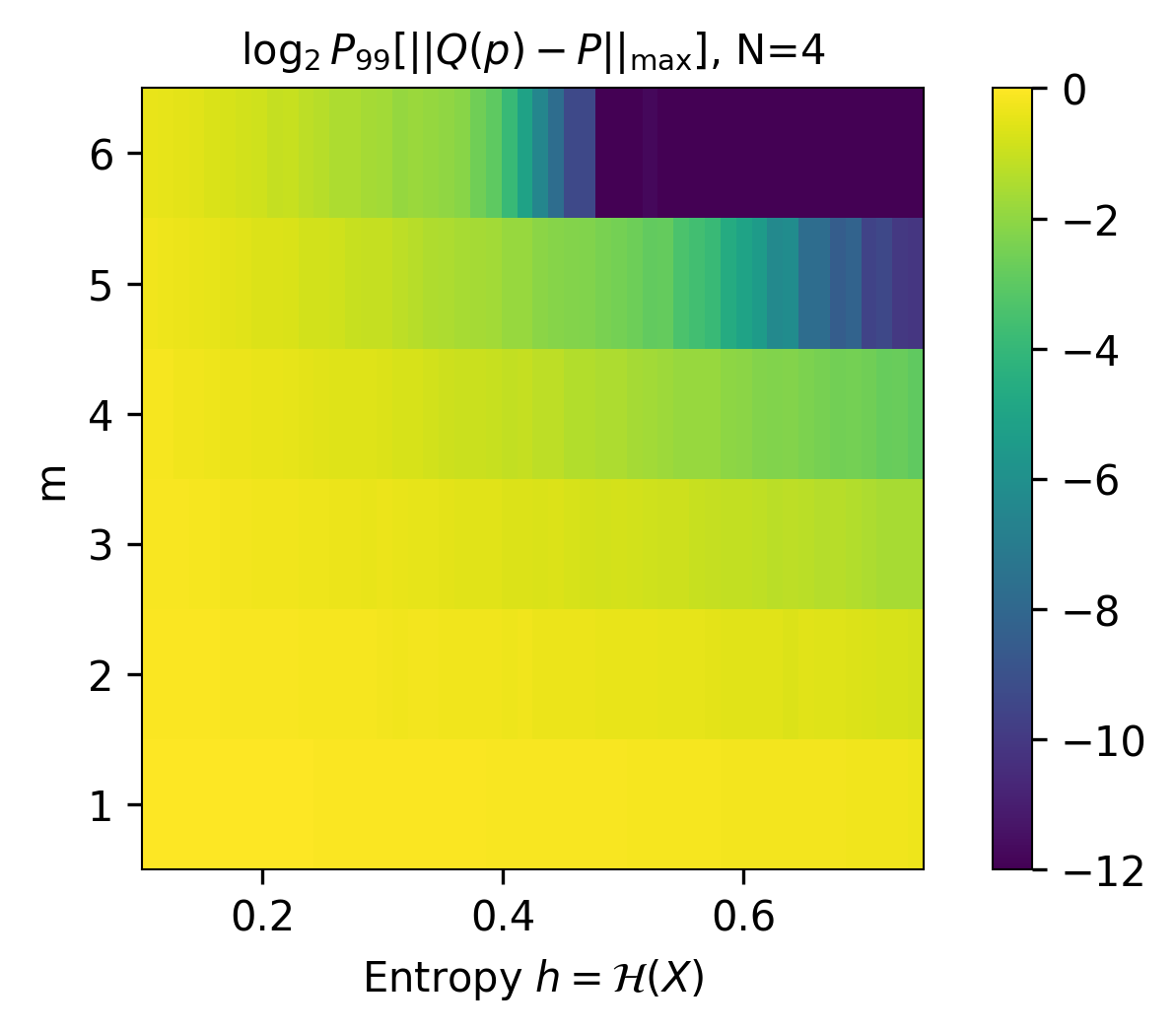}
\label{fig:p99_embed_N4}
\end{subfigure}	
\begin{subfigure}{0.33\textwidth}
\includegraphics[width=\linewidth]{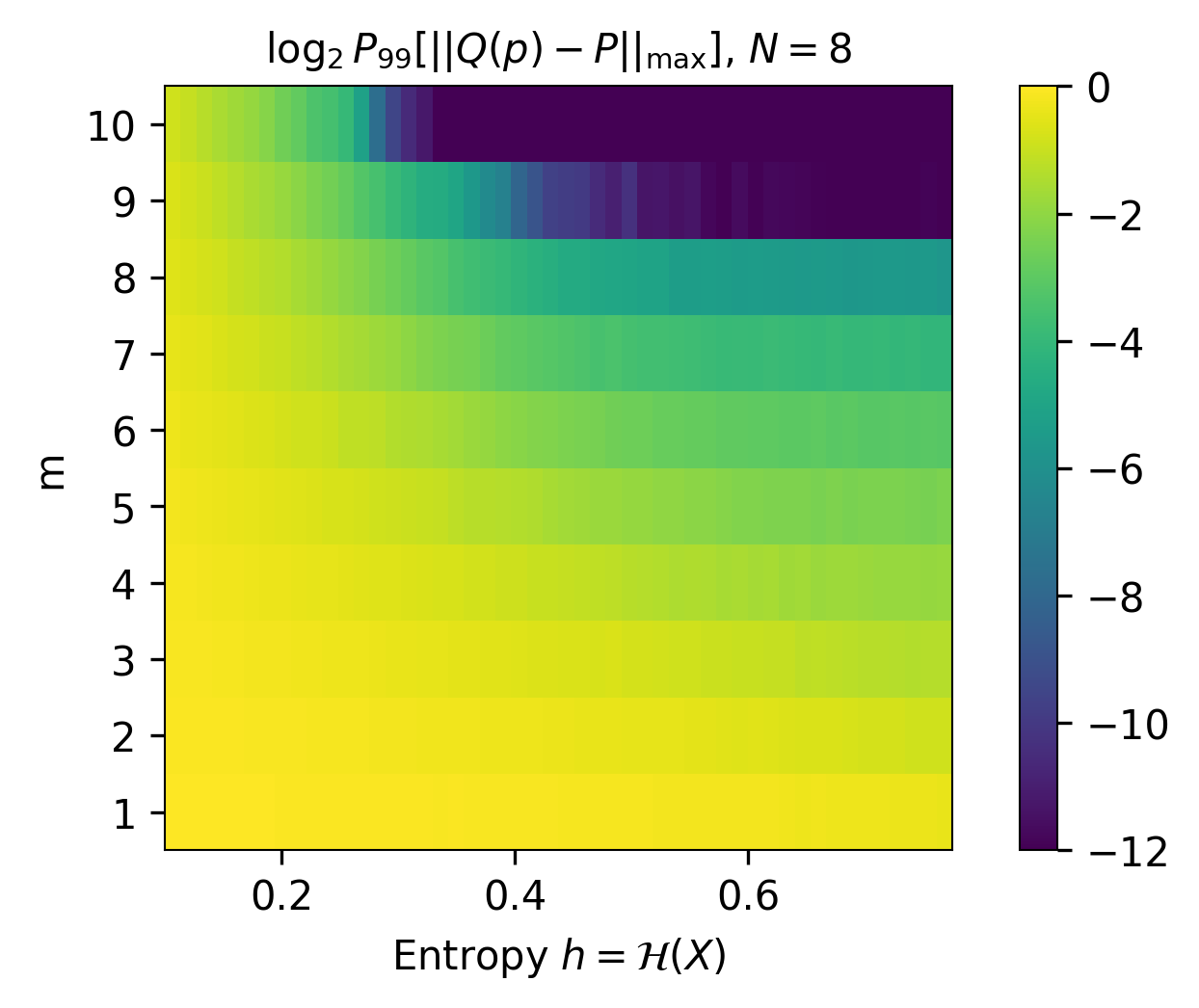}
\label{fig:p99_embed_N8}
\end{subfigure}	
\begin{subfigure}{0.33\textwidth}
\includegraphics[width=\linewidth]{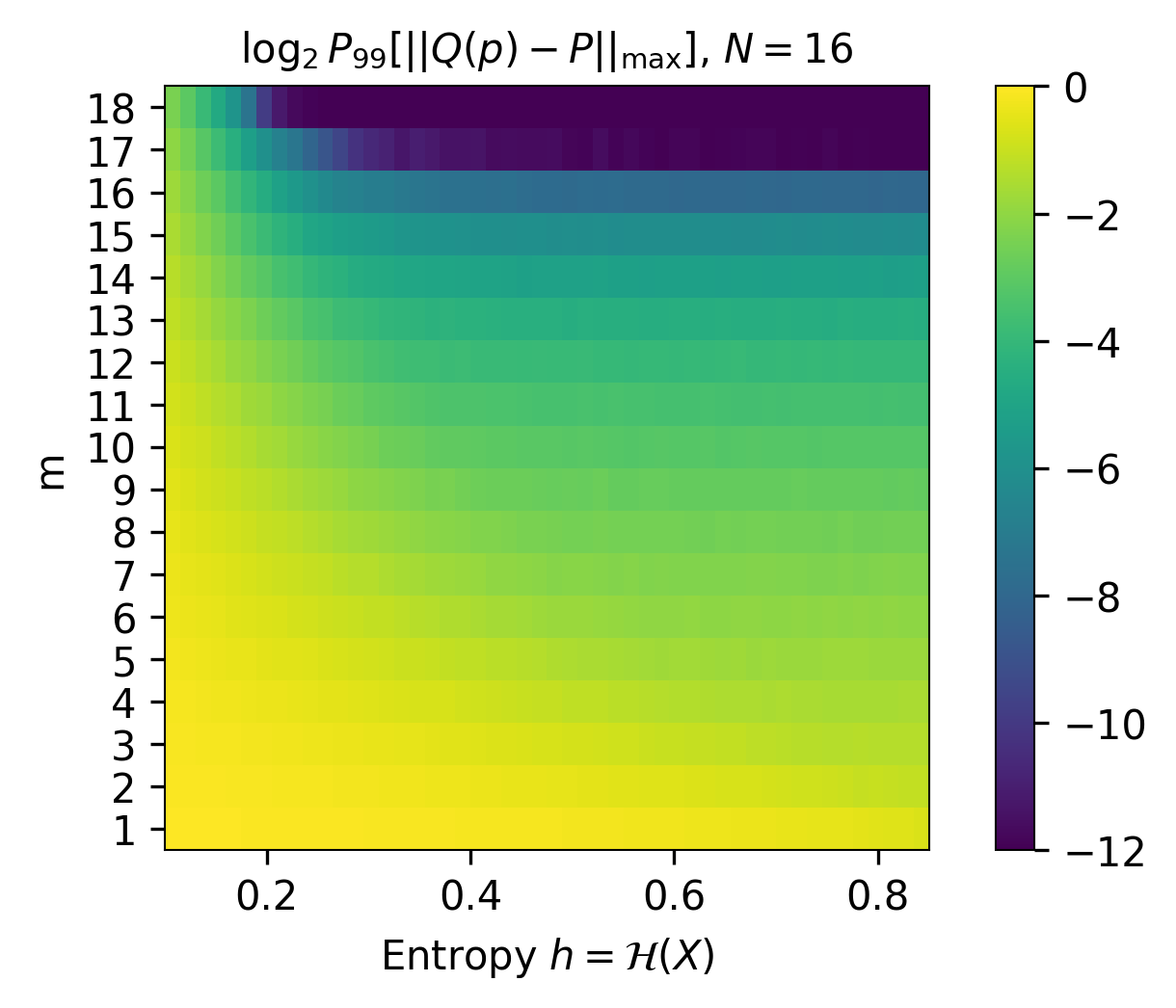}
\label{fig:p99_embed_N16}
\end{subfigure}	
\caption{The 99th percentile of the maximum error, $\log_2 P_{99} [\norm{Q(\bm{p}) - P}_\mathrm{max}]$, in the synthesized approximate linear converter for $N=4$, $N=8$, and $N=16$. For each layer number $m$ and entropy $h$, 512 optimization trials are conducted, and the 99th percentile of the maximum error is estimated using the GEV distribution.}
\label{fig:p99_comp}
\end{figure}

We evaluate the approximation ability of the given linear converter by estimating the distribution of $\norm{Q(\bm{p}) - P}_\mathrm{max}$ and calculating its 99th percentile.
Assuming that errors in each matrix element are independent and identically distributed (i.i.d.), we apply extreme value statistics theory.
This assumption is reasonable because the matrix to be realized is embedded in a unitary matrix, with no special basis vectors. The maximum of a sequence of i.i.d. variables follows the Generalized Extreme Value (GEV) distribution, which is widely applied in fields such as risk analysis and rare event modeling \cite{Fisher1928,Gnedenko1943}.
For evaluation, we consider the sequence of errors in matrix elements, where $\norm{Q(\bm{p}) - P}_\mathrm{max}$ is expected to follow the GEV distribution thanks to the i.i.d. assumption across $N^2$ matrix elements in $Q(\bm{p}) - P$. To estimate this distribution, we run 512 optimization trials varying the initial parameters, with each trial yielding one sample of $\norm{Q(\bm{p}) - P}_\mathrm{max}$.
The distribution is then estimated using maximum likelihood estimation (MLE) based on the 512 samples, with its initial parameters derived via the probability-weighted moments method \cite{Hosking1985}. From the fitted distribution, the 99th percentile of the maximum error is calculated.
In each optimization trial, the target matrix $P$ and the mode mixers $A_i$ are randomly sampled.
To ensure the validity of fitting the GEV distribution and its 99th percentile estimation, we assess the goodness-of-fit using the Kolmogorov-Smirnov test \cite{Kolmogorov1933,Smirnov1948}. In all cases where the converter functions as an approximate converter, the null hypothesis that the observed samples originate from the GEV distribution cannot be rejected at the 5\% significance level. In contrast, when the converter is exactly configured, the null hypothesis is rejected. This is because the assumption of i.i.d. no longer holds in exact configuration, as the error becomes limited by numerical precision during the optimization.

\subsection{Results}
Figure \ref{fig:p99_comp} shows the 99th percentile of the maximum norm, $P_{99}[\norm{Q(\bm{p})-P}_\mathrm{max}]$, as a function of the number of layers $m$ and the entropy of matrix $\mathcal{H}(X)$, for $N=4$, $N=8$, and $N=16$. For $m \geq N+1$, the norm decreases significantly, consistent with the results shown in Figure \ref{fig:block_encoding_layer_vs_error}.
Similar to the result in Fig. \ref{fig:entropy_vs_frobenius_by_unitary}, the reduction of the maximum norm saturates when the entropy exceeds a certain threshold.
For $m < N+1$, where the number of layers is insufficient, the results indicate that the converter operates as an approximate converter, achieving matrices with errors dependent on both the number of layers and the entropy of the mixers.
This implies that the converter can serve as an approximate converter within a defined maximum tolerable error. Increasing $m$ leads to a reduction in the 99th percentile of the maximum norm for a fixed entropy.
Comparing different $N$, fewer layers $m$ are required to achieve the same accuracy for larger $N$, highlighting the scaling advantage of the approximate converter.

\begin{figure}[hbtp]
\centerline{\includegraphics[width=75mm]{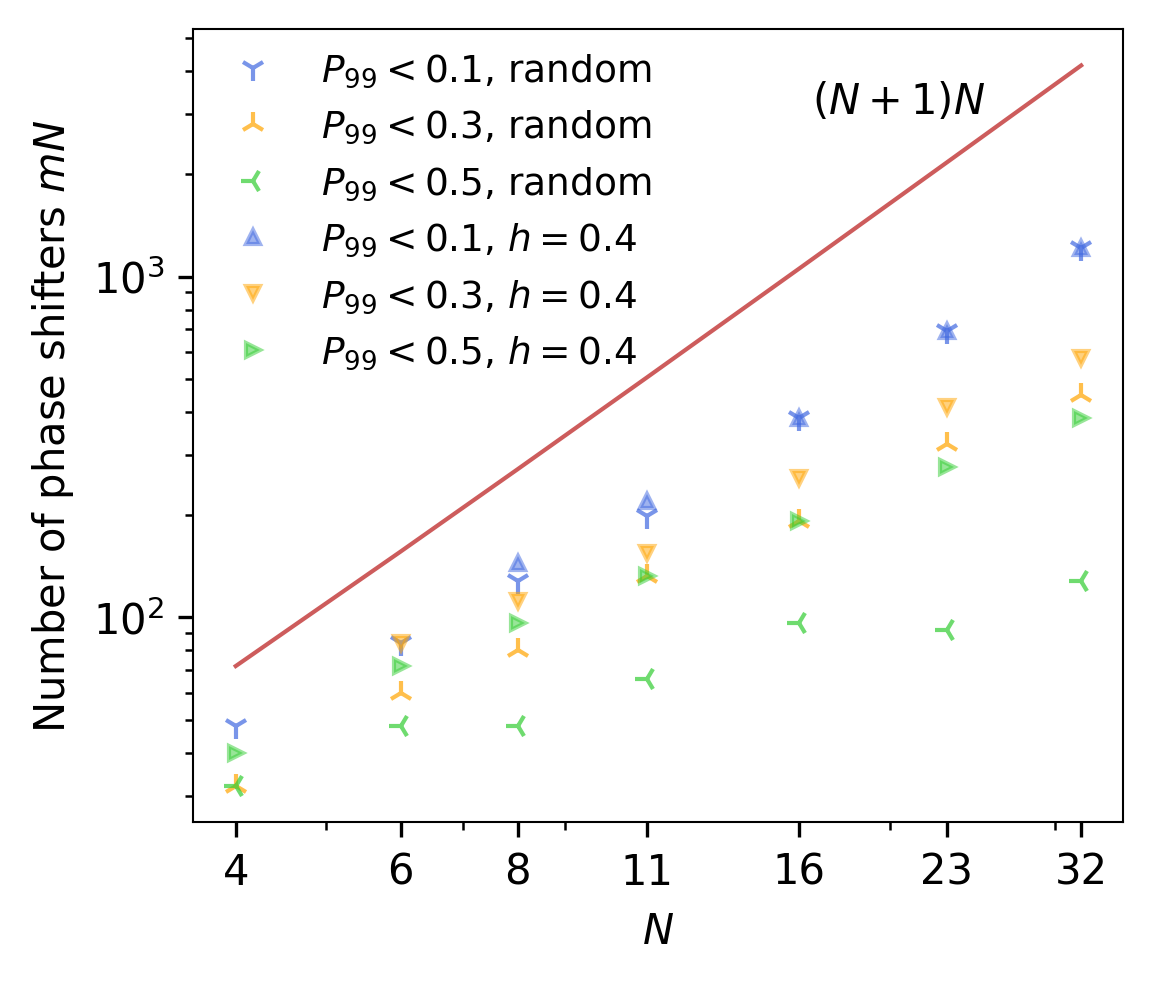}}
\caption{Scaling of the number of phase shifters for the exact and approximate converters. The red solid line shows the case of $m=N+1$, corresponding to the exact synthesis.}
\label{fig:scaling_p99}
\end{figure}

To further explore the scaling properties of the approximate converter, we determined the minimum number of phase shifters, $mN$, required to achieve a specified upper bound on the maximum norm. Figure \ref{fig:scaling_p99} illustrates the least number of phase shifters $mN$ needed to keep $P_{99}$ below a given tolerable error. The red solid line corresponds to $m = N+1$, representing exact synthesis of the linear converter, while markers denote the approximate converter.
For each error bound, we evaluated approximate converters with mixers randomly sampled from the Haar measure, as well as converters using low-entropy mixers ($h=0.4$).
The approximate converter requires significantly fewer phase shifters than the exact converter. For $N=32$, the number of phase shifters is reduced by more than 95\% when allowing an error of $P_{99} < 0.5$.
The use of low-entropy mixers increases the required number of phase shifters compared to the Haar-random case, though the difference becomes smaller as the error tolerance decreases. In some cases, low-entropy mixers with small $N$ cannot satisfy the given error bounds, specifically for $P_{99} < 0.1$ when $N=4$ or $N=6$, and for $P_{99}<0.3$ when $N=4$.
Note that the figure is presented on a double-logarithmic scale, with exact synthesis requiring $mN = (N+1)N = O(N^2)$ phase shifters. In contrast, the approximate converter exhibits sub-quadratic scaling. Fitting a function $bN^\beta$ to $mN$ provides scaling parameter $\beta$, and the scaling is estimated for each case. Given error bounds of $P_{99}<0.1, 0.3$, and $0.5$, converters with Haar randomly sampled mixers show scaling of $\beta=1.66, 1.20$, and $0.61$, respectively. With low-entropy mixers, converters show scaling of $\beta=1.63, 1.18$, and $1.01$ for the same error bounds.
As the error tolerance increases, the scaling of the approximate converter becomes more gradual, demonstrating its scaling advantage.

\section{Model Quantization Applied to the Approximate Converter}
\label{sec:model_quantization}
Quantization of machine learning models compresses weights so that they are represented with a few bits while maintaining the model's accuracy. This enables models to be executed on semiconductor platforms with low computing precision, helping to reduce required computing power and memory.
In this section, we apply the quantization technique to a model that is executed on the proposed optical approximate converter, and demonstrate that a text classification task can be performed without degrading output accuracy.
We train an long short-term memory (LSTM) \cite{Sepp1997} model for a text classification task on the AG's news dataset \cite{AGNews,ZhangNIPS2015} and numerically simulate its execution on the proposed approximate converter. Weight quantization and the Straight-Through Estimator (STE)\cite{Bengio2013} are used during training to make the model weights robust against errors in computation. Our results show that the quantization technique prevents degradation in the model's outputs, indicating that quantization is advantageous not only for semiconductor platforms but also for optical analog computing platforms.

\subsection{Text classification model and its training}
\begin{figure*}[hbtp]
\centerline{\includegraphics[width=130mm]{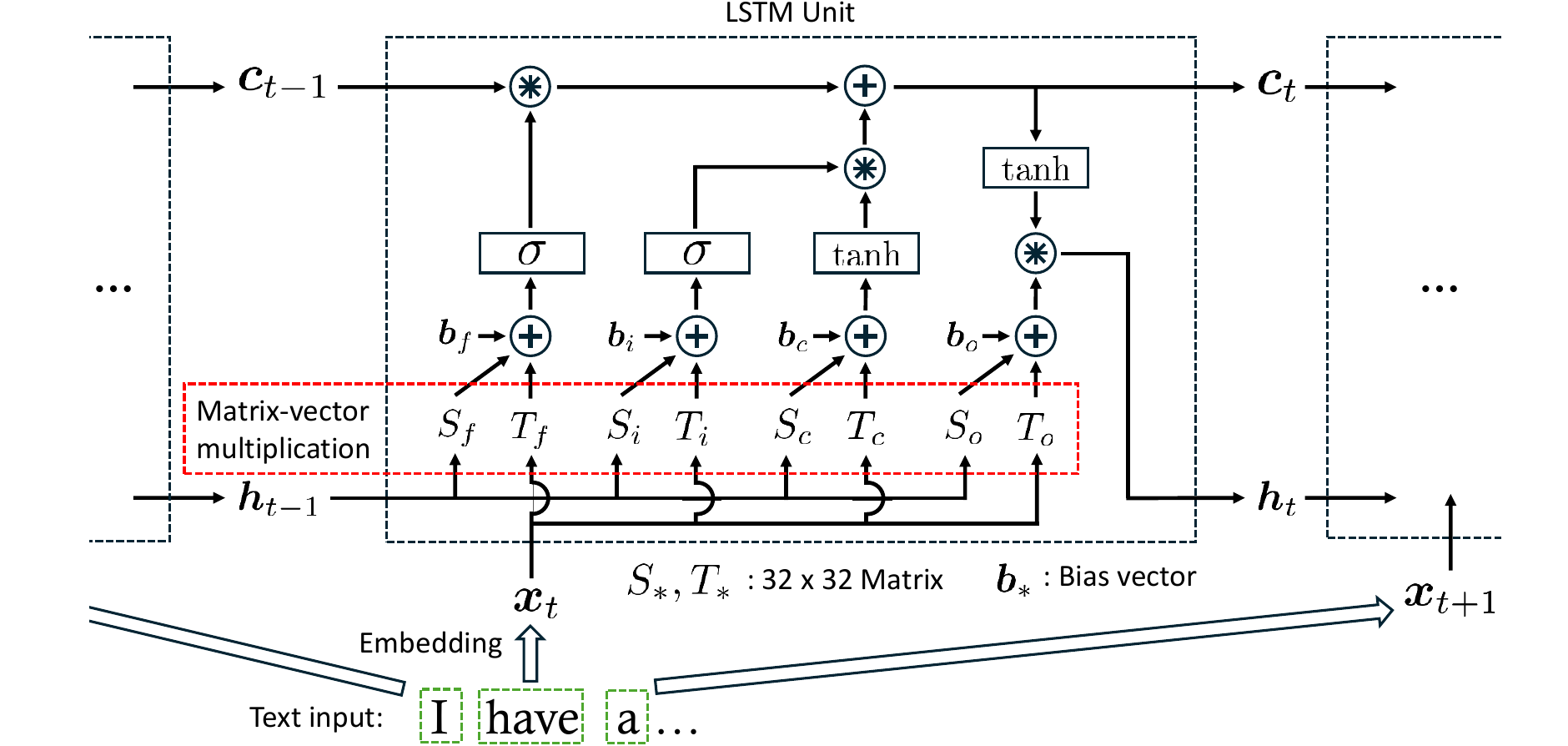}}
\caption{Structure of the LSTM model. Matrix-vector multiplications surrounded by the dashed red line are replaced by the approximate converter after training. Bold letters and solid arrows represent 32-dimensional vectors; $\sigma$ denotes element-wise sigmoid functions, and $\tanh$ denotes element-wise hyperbolic tangent functions. Multiplications between vectors are defined as element-wise (Hadamard) products.}
\label{fig:LSTM}
\end{figure*}

The LSTM model is a recurrent neural network architecture capable of solving tasks that require handling long-term correlations in sequential data. Figure \ref{fig:LSTM} shows the architecture of the model used for text classification. The model takes a sequence of real-valued vectors as input and iteratively updates its internal state vectors. In each iteration, the model performs nonlinear operations and matrix multiplications, updating the state vectors $h$ and $c$.
The matrices, denoted by $S$ and $T$, perform matrix-vector multiplications to map a vector into the same dimensional space during each LSTM iteration. Separate matrices are defined for four nonlinear operation paths, labeled $f, i, c,$ and $o$. The input text to the model is fed through a vector embedding layer, implemented using \texttt{nn.Embedding} in PyTorch\cite{PyTorch}.
We train the model on a GPU to acquire these matrices, and afterward, matrix-vector multiplications are executed using simulated approximate converters. For numerical demonstration, we set the dimension $N=32$.

For the task solved by the LSTM model, we use the AG's news dataset for text classification. The AG's news dataset consists of news articles classified into four categories. We define the task as predicting the category of a news article from its text input. The category prediction is made by applying a linear mapping to the final state vector $\bm{h}$, and the loss function is defined by \texttt{nn.CrossEntropyLoss} in PyTorch. The training set consists of 30,000 samples for each category. For model evaluation, we use 1,000 samples not included in the training set and calculate the accuracy as the ratio of correctly predicted samples out of four categories.

\subsection{Quantization-aware training and model evaluation}
We use weight quantization and the Straight-Through Estimator (STE) during training to enhance the model’s robustness against computational errors. In this context, weights refer to the numerical elements of matrices that determine the model’s transformations. During training, the model first computes its output with matrix elements truncated to discretized values. Specifically, given a matrix element $x$ and bit precision $q$, the value is truncated to $s[x/s]$, where $s = (2^{q} - 1)^{-1}$ is the quantization scale and $[\cdot]$ denotes rounding to the nearest integer. This quantization process deliberately introduces errors into the computation.
To compute gradients, the STE method bypasses the quantization during backpropagation, allowing the gradient to be calculated as if no quantization had occurred. By training the model in the presence of quantization, the matrix elements are optimized to make the entire model robust to computational errors. In our numerical experiments, we trained three models: one normal model without quantization, and two models trained with quantization and STE at $q=4$ and $q=8$, respectively. We trained the models for 5 epochs using the Adam optimizer with a learning rate of $0.001$.

After training the LSTM model with quantization, we replace the matrix-vector multiplications in the model with multiplications performed on the approximate converter. Each matrix in the model is approximated using a converter with insufficient layers, as described in Section \ref{sec:approx_converter}. All mixers in the converter are sampled from the Haar measure. Before approximation, matrices are scaled so that their maximum singular value is less than 1, and the results are scaled back after matrix-vector multiplication.
Because the approximated matrices are complex-valued, we take the real part of the resulting vectors after multiplication, which corresponds to homodyne detection for each mode in the optical device. To evaluate the model, we vary the number of layers $m$ in  the approximate converter and examine the impact of approximation on the model’s classification accuracy.

\subsection{Results}
\begin{figure}[hbtp]
\centerline{\includegraphics[width=75mm]{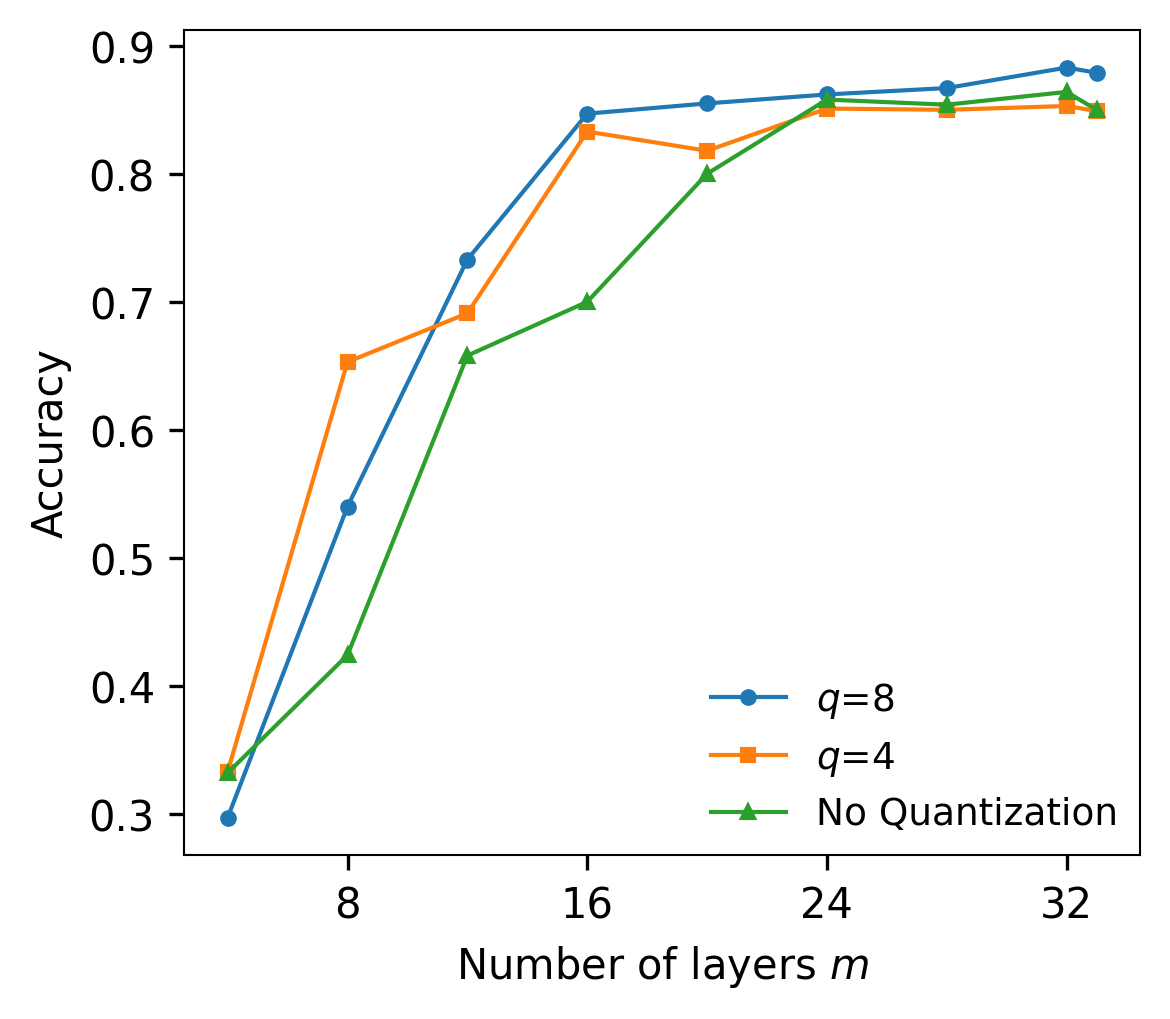}}
\caption{Accuracy of text classification when matrix multiplications in the model are performed on the approximate converters, while varying the number of layers $m$.}
\label{fig:acc_comp_quantization}
\end{figure}

Figure \ref{fig:acc_comp_quantization} shows the model accuracy as a function of the number of layers $m$ in the approximate converter. Cases with $m=4, 8, \ldots, 32,$ and $33$ are investigated. Note that $m=33$ corresponds to the exact synthesis case. All models exhibit a drop in accuracy as the number of layers is reduced. The model without quantization maintains its accuracy for $m \geq 24$, whereas the quantized models maintain their accuracy for $m \geq 16$. For the $q=8$ case, the accuracy drop at $m=16$ is 3.2\% compared to $m=32$. Quantization with $q=8$ improves accuracy by 14.7\% at $m=16$ compared to the non-quantized model.

\section{Conclusion}
\label{sec:conclusion}
We proposed an approximate linear converter based on MPLC for realizing a configurable and compact matrix-vector multiplication device. Our numerical results demonstrate that permitting errors in the converter's synthesis enables sub-quadratic scaling in the number of phase shifters, a significant improvement over the $O(N^2)$ scaling required for exact synthesis. We introduced Shannon matrix entropy as a metric for mode mixing capability and showed that the entropy of mixers can be reduced without compromising the universality. Additionally, we analyzed the structure of general linear converters composed of unitary converters and numerically demonstrated that the BE scheme outperforms the SVD scheme in iterative configuration. 
Through a text classification task, we demonstrated that the approximate converter can reduce the number of required phase shifters for matrix-vector multiplication while maintaining model accuracy. Model quantization was found to be a promising approach for deploying models on approximate computing platforms.
We believe this approach will facilitate scalable implementations of optical linear converters and advance the field of approximate computing using optics.

\appendix
\section{Degrees of freedom of the matrix root in Eq. \ref{eq:block_encoding_embedding}}
\label{sec:proof_dof_root}
We present a proof for the unitary degrees of freedom of $\sqrt{I - Q Q^\dagger}$ in Eq. \ref{eq:block_encoding_embedding}, where $Q$ is a general matrix with singular value is less than or equal to 1.
This proof also applies to $\sqrt{I-Q^\dagger Q}$.
First, we show that $I - Q Q^\dagger$ is a positive semi-definite matrix. The singular value decomposition of $Q$ is given by $Q = V_1 D V_2$, where $V_1$ and $V_2$ are unitary matrices, and $D$ is a diagonal matrix with diagonal elements $0 \leq \sigma_i \leq 1$.
The term inside the square root can be expressed as $I- Q Q^\dagger = I - (V_1 D V_2)(V_2^\dagger D V_1^\dagger) = V_1 (I - D^2) V_1^\dagger$.
Since $0 \leq \sigma_i \leq 1$, all diagonal elements of $I - D^2$ are non-negative. As a result, the matrix $V_1 (I - D^2) V_1^\dagger$ = $V_1 (I - D^2) {V_1}^{-1}$ represents a diagonalization with eigenvalues that are all greater than or equal to zero, meaning that the matrix is positive semi-definite. Since the square-root of a positive semi-definite matrix has unitary degree of freedom \cite{Antoniou2021}, we conclude that $\sqrt{I - Q Q^\dagger}$ has unitary degrees of freedom.


\bibliography{sample}

\end{document}